\documentclass[letterpaper,twocolumn,preprintnumbers,nofootinbib]{revtex4-1}
\usepackage{amsmath,amssymb,graphicx}
\usepackage{xcolor}
\newcommand{\I}{\mathrm{i}}
\newcommand{\E}{\mathrm{e}}
\newcommand{\D}{\mathrm{d}}

\usepackage[normalem]{ulem}

\begin{document}

\title{Resonant Loop Interferometers for High-Frequency Gravitational Waves}

\author{Jan Heisig}
\email[E-mail: ]{heisig@physik.rwth-aachen.de}
\thanks{\\ ORCID: \href{https://orcid.org/0000-0002-7824-0384}{0000-0002-7824-0384}.}
\affiliation{Institute for Theoretical Particle Physics and Cosmology, RWTH Aachen University, 52056 Aachen, Germany}

\begin{abstract}
Gravitational waves at kilohertz and higher frequencies offer a unique probe of the early Universe at temperatures well beyond the reach of the cosmic microwave background, corresponding to energy scales $\gtrsim 10^9\,$GeV. Existing detector concepts fall many orders of magnitude short of the big-bang nucleosynthesis (BBN) bound on the stochastic background in this regime. We propose a new interferometric architecture based on closed optical loops, in which the gravitational-wave–induced phase shift accumulates coherently over many traversals. This produces sharp, narrowband resonances whose predictable comb structure provides a distinct experimental signature. For a folded loop with parameters compatible with the Einstein Telescope infrastructure, and finesse values of order $500$, we project sensitivity that approaches and even surpasses the BBN bound up to tens of kilohertz after one year of integration. Such loop interferometers thus open a realistic and distinctive path toward exploring high-frequency stochastic gravitational-wave backgrounds.

\end{abstract}

\maketitle

\section{Introduction}

Gravitational waves (GWs) provide a unique observational window into the early Universe, reaching epochs far earlier than the cosmic microwave background (CMB). A stochastic gravitational-wave background (SGWB) encodes information about high-energy processes in the primordial plasma, including phenomena far beyond the reach of laboratory experiments. In particular, features at frequencies $f \gtrsim \mathrm{kHz}$ correspond to cosmic temperatures above $10^9\,\mathrm{GeV}$, and can thus probe physics at scales associated with grand unification and other extensions of the Standard Model.

Efforts to access this high-frequency regime are ongoing, ranging from resonant electromagnetic detectors to mechanical systems and novel interferometric concepts. These approaches face formidable challenges: thermal and quantum noise become increasingly severe at high frequency, and practical arm lengths are limited, leading to rapidly degrading sensitivity. As summarized in recent reviews~\cite{Aggarwal:2020olq,Aggarwal:2025noe}, proposed instruments remain many orders of magnitude away from the sensitivity required to probe SGWB amplitudes at the level of the big-bang nucleosynthesis (BBN) bound.

Recent work has explored modified interferometric geometries that fold or kink the optical path in order to enhance high-frequency sensitivity, including L-shaped resonant configurations and hybrid designs that combine folded paths with Michelson-like readout schemes~\cite{Zhang:2022yab,Guo:2024cgz}. These studies demonstrate that departures from conventional Fabry--P\'erot Michelson layouts can open new opportunities in the kilohertz regime, motivating further exploration of alternative geometrical principles for GW detection.

In this work, we put forward a complementary strategy: the coherent accumulation of the GW-induced phase shift in closed interferometric loops, probed through counter-propagating optical beams. The central observation is that, due to the transverse-traceless nature of GWs, changes in the propagation direction of a laser beam can convert the oscillatory GW strain into a net phase accumulation. As clockwise (CW) and counter-clockwise (CCW) beams traverse a closed path with multiple direction changes, their projections onto the GW strain alternate in sign. At specific wavelengths determined by the loop geometry, an optical wavefront propagating around the loop can sample spacetime in a synchronized manner, consistently experiencing either a stretch or a squeeze along each segment. Under these conditions, the GW-induced phase shifts add coherently over successive traversals, leading to resonant enhancement at discrete frequencies.

While the resulting response is intrinsically narrowband, this selectivity is a key advantage rather than a limitation. The predictable resonance pattern provides a distinctive, self-identifying spectral signature that is difficult to mimic by instrumental noise or environmental disturbances. When implemented in a differential readout between CW and CCW propagation, the resonant GW signal appears exclusively in the differential channel, furnishing a particularly clean hallmark of a genuine GW origin. Such closed-loop geometries can, in principle, be embedded within terrestrial facilities, for example within the underground infrastructure foreseen for the Einstein Telescope (ET), or deployed in space-based settings.

The purpose of this work is to establish the physical principle, response properties, and parametric sensitivity reach of resonant loop interferometers, with particular emphasis on folded-loop realizations compatible with terrestrial operation. Detailed optical designs, noise budgets, and optimization strategies are left to future dedicated studies.

This paper is organized as follows. In Sec.~\ref{sec:phase_accumulation} we introduce the phase-accumulation mechanism in closed optical paths and develop a general formalism for GW-induced phase shifts. Sec.~\ref{sec:realistic_implementations} discusses realistic folded-loop implementations and the suppression of rotational non-reciprocity. In Sec.~\ref{sec:sensitivityfeasibility} we analyze the detector response and present sensitivity projections calibrated to ET performance. We conclude in Sec.~\ref{sec:conlcusion}. Additional technical details are provided in the Appendices.

\section{Concept and principle}
\label{sec:phase_accumulation}

\subsection{Coherent phase accumulation via direction changes}
\label{sec:physical_mechanism}

Gravitational waves induce optical phase shifts through the tidal deformation of spacetime along the photon trajectory. In transverse--traceless gauge, this effect depends on the projection of the GW strain tensor onto the instantaneous direction of light propagation. As a consequence, the accumulated phase shift is sensitive not only to the length of the optical path, but also to changes in its direction.

This feature is exploited most transparently in a closed optical loop composed of straight segments. As the laser beam changes direction at each corner, its projection onto the GW strain alternates in sign. This follows from the transverse-traceless structure of the metric perturbation and underlies the familiar quadrupolar deformation pattern of GWs. When the oscillation of the GW is synchronized with the light propagation around the loop, these alternating contributions do not cancel but instead add coherently.

\begin{figure}[t]
\centering
\includegraphics[width=0.49\textwidth]{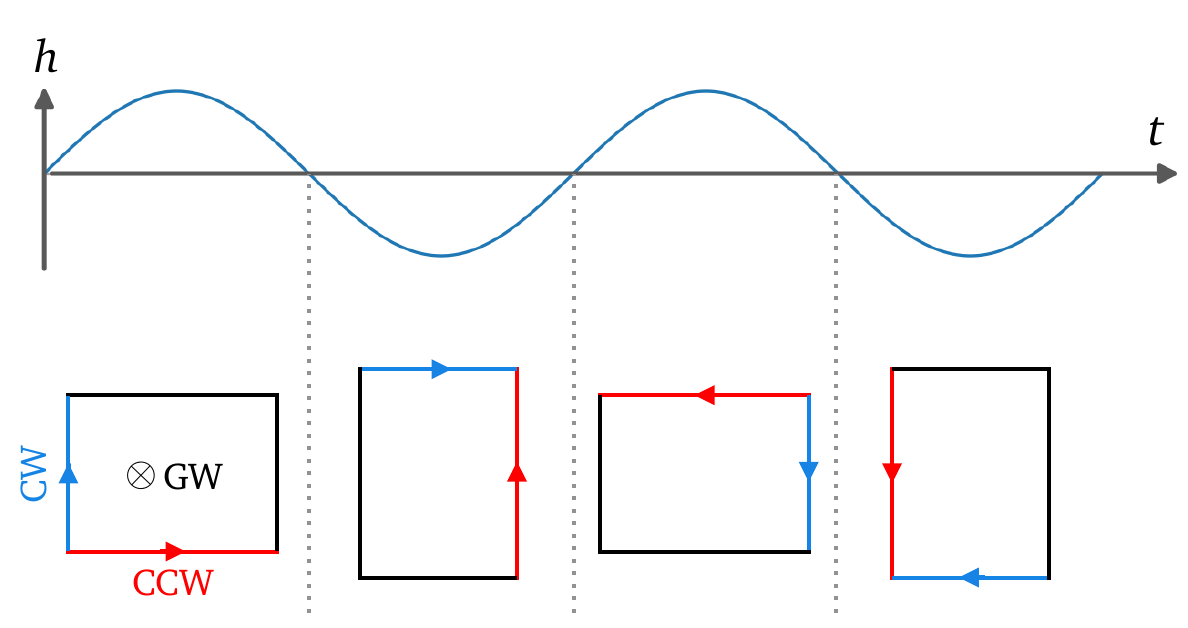}
\caption{Illustration of the detection principle for a square-loop resonator under a normally incident GW with $\lambda_{\rm GW}=2L$. 
Top: two cycles of the GW strain $h(t)$. 
Bottom: four snapshots over one round trip showing the quadrupolar deformation of the loop; blue/red arrows indicate CW and CCW propagation. 
Alternating stretches and squeezes yield four $\pi$-shifted projections that add coherently at the resonance condition.}
\label{fig:square_loop_principle}
\end{figure}

Figure~\ref{fig:square_loop_principle} illustrates this mechanism for a square loop under normal incidence of a GW\@. During one round trip, successive segments sample alternating stretches and squeezes of spacetime. At specific wavelengths set by the loop geometry, an optical wavefront propagating around the loop can sample spacetime in a synchronized manner, always experiencing a stretch (or squeeze) on each segment of the path. In this case, the GW-induced phase shifts add coherently along the entire loop, leading to a resonant enhancement of the accumulated optical phase. 

For the square loop under normal incidence, clockwise (CW) and counter-clockwise (CCW) propagation lead to GW-induced phase shifts of opposite sign. This reflects the dependence of the accumulated phase on the orientation of the optical path and permits the construction of differential observables between counter-propagating beams, an aspect that will play an important role in our discussion below.

\subsection{General formulation for piecewise-straight optical paths}
\label{sec:general_formalism}

We now formulate the GW-induced optical phase shift for a general closed path composed of straight segments. This provides a unified framework applicable to all geometries considered in this work.

For a photon propagating along an optical path $\mathcal{P}$, the GW-induced phase shift accumulated over one traversal is given, to first order in the metric perturbation, by
\begin{equation}
\Delta\Phi
\;=\;
\frac{2\pi}{\lambda_{\rm laser}} \,
\frac{1}{2}
\int_{\mathcal{P}}
h_{ij}(t,\mathbf{x})\, \hat e^{\,i}\hat e^{\,j}\, \D x ,
\label{eq:path_integral}
\end{equation}
where $\hat e$ denotes the local unit tangent vector to the photon trajectory, and the metric perturbation is evaluated along the null path, with the GW taken at the retarded spacetime point of the photon.

We consider an optical path composed of $N_{\rm seg}$ straight segments, each characterized by a unit tangent $\hat e_s$, length $L_s$, and starting point $\mathbf{x}_{s,0}$. For a plane GW propagating in direction $\hat k$ with angular frequency $\omega_{\rm GW}$, the GW phase varies linearly along each segment, allowing the integral in Eq.~\eqref{eq:path_integral} to be performed analytically.

The phase shift accumulated on segment $s$ during the $m$-th traversal can be written in the compact form
\begin{equation}
\begin{split}
\!\!\!\!\!\delta\Phi^{(m)}_s
=
\frac{2\pi}{\lambda_{\rm laser}}
\frac{L_s}{2}
\Bigg[
h_{+} P^{+}_s
\cos\!\left(\beta^{(m)}_s + \frac{\alpha_s L_s}{2}\right)\;\,\qquad
\\
\quad\,\qquad -\,
h_{\times}P^{\times}_s
\sin\!\left(\beta^{(m)}_s + \frac{\alpha_s L_s}{2}\right)\!
\Bigg]\,
\mathrm{sinc}\!\left(\frac{\alpha_s L_s}{2}\right),
\label{eq:segment_phase}
\end{split}
\end{equation}
where $P^{+,\times}_s$ denote the polarization projection factors,
$\alpha_s = \omega_{\rm GW}(1-\hat k\!\cdot\!\hat e_s)$,
and $\beta^{(m)}_s$ encodes the segment position within the loop and the traversal index. Explicit definitions are collected in Appendix~\ref{app:closed_form_sums}. 

The total phase accumulated after $n$ completed traversals of the closed path $\mathcal{P}$ is obtained by summing Eq.~\eqref{eq:segment_phase} over all segments and all round trips. We define
\begin{equation}
\Delta\Phi(n)
\equiv
\sum_{m=0}^{n-1}
\sum_{s\in\mathcal{P}}
\delta\Phi_s^{(m)} .
\label{eq:DeltaPhi_n}
\end{equation}
For the geometries considered in this work, the sums can be evaluated in closed form. These expressions are convenient for numerical implementations (as used in the sensitivity calculations presented below) and are displayed in Appendix~\ref{app:closed_form_sums}. Note that here and in the following, we assume that the GW is present and phase-coherent over the duration of the optical propagation, i.e. over many loop traversals. This is appropriate for long-lived or stochastic GW backgrounds, which are the focus of this work.

\subsection{Worked example: square loop at normal incidence}
\label{sec:square_loop_example}

To make contact with the physical picture discussed in Sec.~\ref{sec:physical_mechanism}, we evaluate the general expression above for the pedagogical case of a square loop of side length $L$ lying in the $xy$-plane, with a GW propagating along the $z$-direction.

In this configuration, the GW strain is purely transverse, and for a plus polarization aligned with the $x$ and $y$ axes, the metric perturbation satisfies $h_{xx} = -h_{yy}$. The four straight segments of the loop therefore acquire alternating projection factors $P^{+}=\pm1$, while the cross polarization does not contribute.

Evaluating the segment-level phase shifts and summing over one completed loop, the total GW-induced phase accumulated after $n$ round trips can be written in closed form. At the discrete resonance wavelengths
\begin{equation}
\lambda_{\rm GW}
=
\frac{2L}{2\ell+1},
\qquad
\ell = 0,1,2,\ldots ,
\label{eq:square_resonance}
\end{equation}
the phase advance of the GW over one loop traversal equals an odd multiple of $\pi$. Under this condition, the GW-induced phase shifts add constructively over successive traversals and the accumulated phase grows linearly with the number of traversals, $n$, as well as with the geometric length scale of the interferometer. Specifically, one finds
\begin{equation}
\Delta\Phi_{\rm res}(n)
=
\frac{8\,h_0\,n\,L}
{\lambda_{\rm laser}\,(2\ell+1)}
\sin\phi_0 ,
\label{eq:square_resonant_phase}
\end{equation}
where $h_0$ denotes the GW strain amplitude and $\phi_0$ the initial GW phase. The explicit time dependence of the observed signal follows from $\phi_0 \to \phi_0 + \omega_{\rm GW} t$.

Equation~\eqref{eq:square_resonant_phase} provides a quantitative expression of the coherent phase-accumulation mechanism illustrated in Fig.~\ref{fig:square_loop_principle}. Away from the resonance condition, the alternating projections lead to partial cancellation, resulting in a strongly frequency-selective response.

\section{Realistic implementations}
\label{sec:realistic_implementations}

The phase-accumulation mechanism described in Sec.~\ref{sec:phase_accumulation} was formulated for an optical field traversing a prescribed closed path a fixed number of times in the presence of a GW\@. In practice, achieving a large number of coherent loop traversals requires storing the optical field in a resonant cavity. We therefore turn to the conceptual requirements and constraints associated with implementing this mechanism in resonant loop interferometers.

\subsection{Resonant enhancement in closed optical loops}

A natural way to realize many successive traversals of the same closed optical path is to form a resonant loop cavity. In practice, such a cavity must be actively stabilized so that its total optical path length remains resonant to sub-wavelength precision, using standard cavity-locking techniques analogous to those employed in Fabry-P\'erot interferometers. In such a resonator, the circulating optical field corresponds to a coherent superposition of contributions from many round trips, with the weighting determined by the mirror reflectivities and optical losses. The effective number of coherent round trips is characterized by the cavity finesse $\mathcal{F}$, with
\begin{equation}
\langle n_{\rm rt} \rangle \simeq \frac{\mathcal{F}}{\pi},
\label{eq:nrt_finesse}
\end{equation}
up to geometry-dependent factors of order unity. In this way, the linear growth of the GW-induced phase shift with the number of traversals derived in Sec.~\ref{sec:phase_accumulation} translates directly into a resonant enhancement of the observable signal.

Conceptually, the resonant cavity implements the repeated traversal assumed in Sec.~\ref{sec:phase_accumulation} by continuously injecting light into the loop and coherently superposing the circulating fields. The GW-induced phase shift accumulated along the loop therefore imprints itself on the steady-state intracavity field, provided that successive round trips remain phase coherent and that the loop forms a reciprocal optical resonator in the absence of a GW signal.

\subsection{Rotational constraints and folded-loop realization}

For large-scale loop geometries operated on Earth, the requirement of reciprocity is nontrivial. Any optical loop enclosing a finite area on a rotating platform experiences non-reciprocal phase shifts between CW and CCW propagation. This effect leads to a splitting of the CW and CCW resonance frequencies, commonly referred to as the Sagnac effect. For a loop of total optical length $L_{\rm rt}$ and area vector $\mathbf{A}$ on a platform rotating with angular velocity $\boldsymbol{\Omega}$, the corresponding frequency splitting is
\begin{equation}
\frac{\Delta \nu}{\nu}
=
\frac{4\,\boldsymbol{\Omega}\!\cdot\!\mathbf{A}}
{c\,L_{\rm rt}} .
\label{eq:sagnac}
\end{equation}

For kilometer-scale loops on Earth, this splitting is many orders of magnitude larger than the linewidth of a high-finesse resonator.\footnote{For an ET-scale square loop with $L_{\rm rt}\simeq 40\,\mathrm{km}$ 
and finesse $\mathcal F\sim 500$, the cavity linewidth is 
$\delta\nu \sim c/(L_{\rm rt}\mathcal F)\approx 15\,\mathrm{Hz}$. 
This corresponds to an indicative requirement 
$\Delta\nu/\nu \lesssim \lambda/(L_{\rm rt}\mathcal F)
\sim 5\times 10^{-14}$ 
(for $\lambda=1064\,\mathrm{nm}$) in order for the CW and CCW modes 
to remain simultaneously resonant. Applying 
Eq.~\eqref{eq:sagnac} to a square loop of side length 
$L=10\,\mathrm{km}$ and assuming no geometric suppression 
($\boldsymbol{\Omega}\!\cdot\!\mathbf{A} \sim \Omega A$) yields a fractional Sagnac splitting 
$\Delta\nu/\nu \sim 2\times 10^{-9}$ 
for $\Omega_\oplus=7.29\times10^{-5}\,\mathrm{rad/s}$, 
exceeding the allowable level by roughly five orders of magnitude.}
As a result, simple ring geometries such as square or circular loops cannot be operated in a regime where the circulating field remains resonant over many round trips: the CW and CCW resonance conditions are strongly separated, and the coherent build-up required for resonant GW phase accumulation is lost. Even attempts to suppress this effect by orienting the loop area vector perpendicular to the Earth’s rotation axis would impose severe geometric constraints and remain insufficient for realistic multi-kilometer implementations. While such geometries are therefore useful as pedagogical examples, they are not suitable for practical high-finesse operation on Earth.

\begin{figure}[t]
\centering
\includegraphics[width=0.45\textwidth]{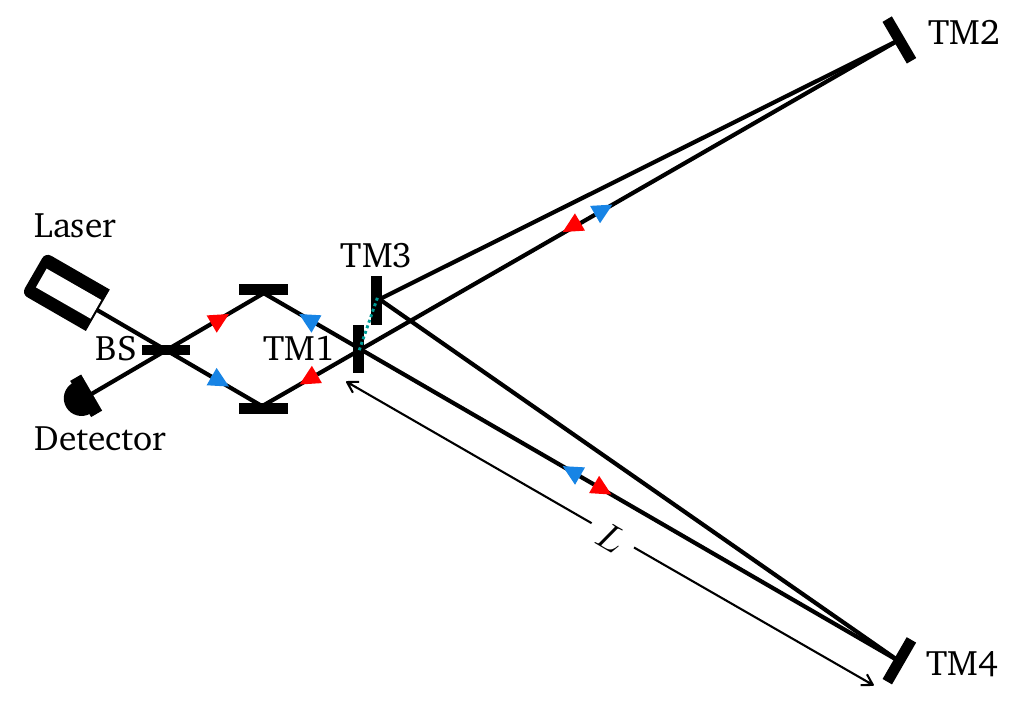}
\caption{Schematic of the folded-loop interferometer consisting of four test masses (TM1--TM4) separated by segments of length $L$ (up to small geometric offsets). The blue and red arrows indicate CW and CCW beam propagation, respectively. The test masses TM1 and TM3 are placed in close proximity and adjusted to lie along a line parallel to the Earth’s rotation axis, suppressing the Sagnac effect.}
  \label{fig:loop_geometry}
\end{figure}

This obstruction can be overcome by modifying the loop geometry such that the Sagnac effect is suppressed to a controllable level. The folded-loop configuration considered here, shown schematically in Fig.~\ref{fig:loop_geometry}, achieves this suppression in two steps. First, the loop is folded such that two nominally distinct mirrors (denoted TM1 and TM3 in Fig.~\ref{fig:loop_geometry}) are brought into close proximity. We denote their displacement vector by $\boldsymbol{\delta} \equiv \mathbf{x}_{\mathrm{TM3}} - \mathbf{x}_{\mathrm{TM1}}$,
with magnitude $|\boldsymbol{\delta}| \ll L$. This eliminates the large enclosed area responsible for the dominant rotational non-reciprocity present in an unfolded loop.

Second, the remaining, parametrically smaller Sagnac effect 
associated with the enclosed area 
$A \sim \mathcal{O}(L|\boldsymbol{\delta}|)$ 
can be further suppressed geometrically. 
By (actively) aligning $\boldsymbol{\delta}$ parallel to the Earth’s rotation axis, $\boldsymbol{\delta}\parallel\boldsymbol{\Omega}_\oplus$, 
the normals of the two slender triangular sub-areas spanned by 
(TM1, TM2, TM3) and (TM1, TM3, TM4) become, to high precision, orthogonal to $\boldsymbol{\Omega}_\oplus$. As a result, the corresponding Sagnac contributions are strongly suppressed, reducing the residual non-reciprocal phase shift to a level compatible with high-finesse operation.

Unlike an unfolded multi-kilometer loop, this alignment does not require orienting long vacuum tunnels parallel to the Earth’s rotation axis; it can be achieved locally through the relative placement and control of nearby mirrors, which we consider feasible within current interferometric alignment tolerances.

Crucially, the folded-loop geometry preserves the GW phase-accumulation mechanism established in Sec.~\ref{sec:phase_accumulation}. The optical path still consists of straight segments joined at finite angles, and the sequence of direction changes responsible for the coherent build-up of the GW-induced phase shift remains intact. At the same time, the suppression of the enclosed area restores reciprocity and allows the formation of a resonant loop cavity compatible with large finesse.

\subsection{Schematic experimental layout and readout considerations}

We now turn to a schematic experimental realization of the folded-loop interferometer, illustrated in Fig.~\ref{fig:loop_geometry}. In this configuration, light is injected into the loop through an input coupler and circulates in both CW and CCW directions (indicated by the blue and red arrows, respectively).\footnote{Here ``CW'' and ``CCW'' serve as convenient labels for the two counter-propagating circulations, even though the intuitive notion of a clockwise sense does not apply unambiguously to the folded-loop geometry.} The circulating fields can be extracted through a common output coupler and directed to a photodetector via a beam splitter (BS), allowing access to their relative phase.

The folded-loop geometry naturally permits the construction of differential observables based on the phase difference between CW and CCW propagation, while common-mode contributions are suppressed by symmetry. At the same time, the resonant enhancement discussed above ensures that the GW-induced phase shift accumulated over many loop traversals is imprinted on the output signal. A detailed discussion of the detector response and the relative role of common-mode and differential readout channels is deferred to Sec.~\ref{sec:response}.

Before closing this section, it is useful to note that the folded-loop geometry gives rise to an analytic resonance condition closely analogous to that of the square loop discussed in Sec.~\ref{sec:phase_accumulation}. For a folded loop composed of four straight segments of length $L$, the GW-induced phase accumulation becomes resonant when the GW wavelength satisfies
\begin{equation}
\lambda_{\rm GW} = \frac{4L}{2\ell+1},
\qquad
\ell = 0,1,2,\ldots .
\label{eq:folded_loop_resonance}
\end{equation}
At these frequencies, successive loop traversals add coherently, leading to a linear growth of the accumulated phase with the number of round trips.
Evaluating the general expressions of Sec.~\ref{sec:general_formalism} at resonance yields a closed-form result for the accumulated phase shift that differs from the square-loop case only by geometry-dependent numerical factors.

\section{Sensitivity and Feasibility}
\label{sec:sensitivityfeasibility}

\subsection{Response function}
\label{sec:response}

We now turn to the response of a resonant loop interferometer to an incident GW\@. The goal of this section is to characterize how the GW-induced phase accumulation derived in Sec.~\ref{sec:phase_accumulation} manifests itself in an experimentally accessible observable, and to define an appropriate response function that captures the intrinsic sensitivity of the detector geometry.

As discussed in Sec.~\ref{sec:realistic_implementations}, in a resonant loop cavity the circulating optical field corresponds to a coherent superposition of contributions that have completed different numbers of round trips. Denoting by $\Delta\Phi(n)$ the GW-induced phase accumulated after $n$ completed traversals of the loop, as defined in Sec.~\ref{sec:general_formalism}, the phase imprinted on the circulating intracavity field can be written schematically as
\begin{equation}
\Delta\Phi_{\rm cav}
=
\sum_{n=0}^{\infty} w_n\, \Delta\Phi(n) ,
\label{eq:linear_superposition}
\end{equation}
where the weights $w_n$ are set by the cavity coupling and losses, see Appendix~\ref{app:closed_form_sums} for more details. For a simple loop cavity with amplitude reflectivity $r$ per round trip (including all losses) one has $w_n \propto r^{n}$, i.e.\ an approximately exponential decay in $n$.

It is often convenient to characterize this decay by an effective number of contributing round trips,
\begin{equation}
\langle n_{\rm rt} \rangle \;\equiv\; \sum_{n=0}^{\infty} w_n,
\label{eq:nrt_def}
\end{equation}
which is related to the finesse via $\langle n_{\rm rt} \rangle \simeq \mathcal{F}/\pi$ up to factors of order unity (cf.\ Eq.~\eqref{eq:nrt_finesse}). The enhancement of the GW-induced phase is governed not only by $\langle n_{\rm rt} \rangle$ but also by the relative phasing of the contributions $\Delta\Phi(n)$. This phasing depends on the GW frequency
through retardation along the loop. When successive traversals add coherently, Eq.~\eqref{eq:linear_superposition} yields an enhancement scaling approximately with $\langle n_{\rm rt} \rangle$; away from coherence the weighted sum leads to partial cancellation.

For the purpose of characterizing the frequency dependence, we consider a monochromatic GW of angular frequency $\omega_{\rm GW}$ with strain
\begin{equation}
h(t) = h_0 \cos(\omega_{\rm GW} t + \phi_0),
\label{eq:monochromatic_gw}
\end{equation}
propagating from direction $(\theta,\varphi)$ with polarization angle $\psi$. To first order in $h_0$, the induced phase shift is linear in the strain and therefore oscillates at the same frequency. We may thus write
\begin{equation}
\Delta\Phi(t;\theta,\varphi,\psi)
=
h_0 \,\Re\!\left[
\widetilde{R}(\theta,\varphi,\psi;\omega_{\rm GW})\,
\E^{\I(\omega_{\rm GW} t + \phi_0)}
\right],
\label{eq:response_complex}
\end{equation}
which defines the (in general complex) response amplitude $\widetilde{R}$. Note that $\widetilde{R}$ should be understood as the linear response amplitude
relating a monochromatic GW drive at frequency $\omega_{\rm GW}$ to the
oscillatory phase response; it does not imply a Fourier transform of an
arbitrary time-domain signal.

For a stochastic background the phase $\phi_0$ is effectively random, and it is natural to characterize the detector by the mean-square response averaged over $\phi_0$, sky direction, and polarization. Averaging over one GW period (equivalently over $\phi_0$) yields
\begin{equation}
\left\langle \Delta\Phi^2 \right\rangle_{\phi_0}
=
\frac{h_0^2}{2}\,
\left|\widetilde{R}(\theta,\varphi,\psi;\omega_{\rm GW})\right|^2 .
\label{eq:phi0_average}
\end{equation}
We therefore define the RMS response to an isotropic background as
\begin{equation}
R_{\rm rms}(\omega_{\rm GW})
=
\sqrt{
\frac{1}{16\pi^2}
\int \D \Omega
\int_{0}^{2\pi} \!\!\D \psi 
\left|\widetilde{R}(\theta,\varphi,\psi;\omega_{\rm GW})\right|^2
}
,
\label{eq:rms_response}
\end{equation}
so that $\langle \Delta\Phi^2 \rangle_{\Omega,\psi,\phi_0} = h_0^2\,R_{\rm rms}^2(\omega_{\rm GW})$.
In Eq.~\eqref{eq:rms_response}, the angular integral extends over the full sky and the polarization angle $\psi$ is assumed to be uniformly distributed.\footnote{For a monochromatic plane wave with fixed direction, the polarization basis can be rotated such that $h_\times=0$, so that averaging over $\psi$ captures the general case.}
This averaged response smooths over directional modulations while retaining the characteristic frequency dependence associated with the geometric resonances of the loop interferometer.

\begin{figure}[t]
\centering
\includegraphics[width=0.48\textwidth]{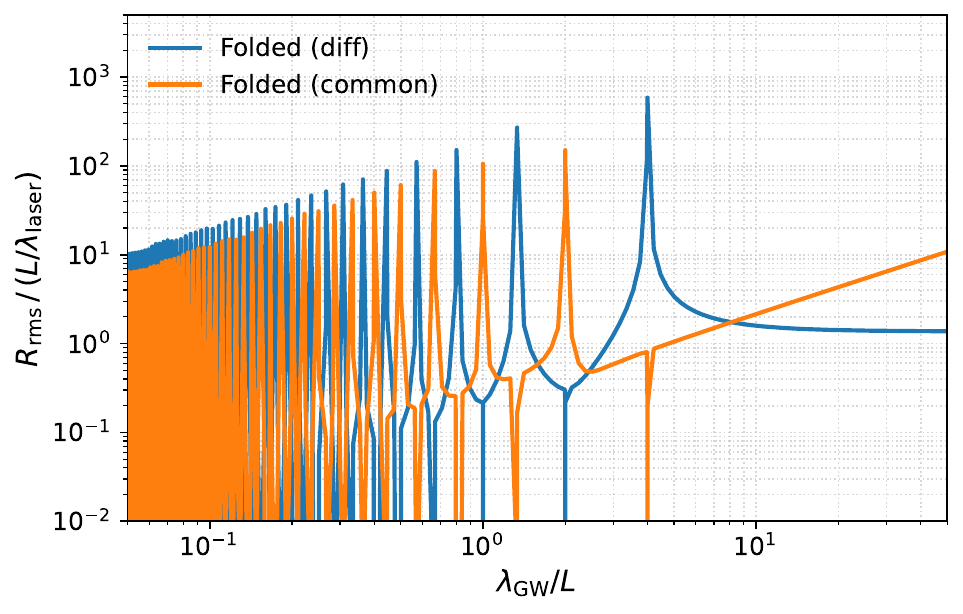}
\caption{Normalized root-mean-square response $R_{\rm rms}$ of the folded-loop interferometer as a function of the GW wavelength in units of the loop length, $\lambda_{\rm GW}/L$, shown for the differential (CW--CCW) and common (CW+CCW) readout modes. The trivial scaling with $L/\lambda_{\rm laser}$ has been factored out, so that the plotted response depends only on the geometry, the number of coherent round trips, and the dimensionless frequency.}
\label{fig:response_folded_loop}
\end{figure}

Figure~\ref{fig:response_folded_loop} shows the resulting RMS response functions as a function of GW frequency for the folded-loop geometry. As a baseline, we consider a configuration where the two legs of the folded loop enclose an angle of $60^\circ$, compatible with the underground infrastructure foreseen for the Einstein Telescope~\cite{Sathyaprakash:2012jk}. We assume a finesse $\mathcal{F}=500$, corresponding to $\langle n_\mathrm{rt}\rangle \simeq160$. A finesse of this level is realistic given present mirror coatings and demonstrated resonator performance~\cite{LIGOScientific:2014pky,Aggarwal:2020umq}. 
The differential response exhibits a sequence of sharp resonances at frequencies corresponding to the analytic condition in Eq.~\eqref{eq:folded_loop_resonance}.\footnote{The response also exhibits secondary features at 
$\lambda_{\rm GW}=2L/(2\ell+1)$, which appear as resonances in the 
common-mode channel. 
These originate from retardation along individual straight segments 
and are distinct from the coherent loop-accumulation resonances at 
$\lambda_{\rm GW}=4L/(2\ell+1)$. Similar retardation features also arise 
in conventional Fabry-P\'erot-Michelson interferometers.} These resonances reflect the coherent accumulation of the GW-induced phase over many loop traversals when the GW wavelength is commensurate with the optical path length. The height of these resonant peaks scales linearly with the effective number of coherent round trips, $\langle n_\mathrm{rt}\rangle$, so that higher cavity finesse leads to a correspondingly stronger enhancement.

For wavelengths above the resonances, $L \ll\lambda_{\rm GW} \ll \langle n_{\rm rt} \rangle L $, the response exhibits distinct baseline behaviors in the two readout channels. The common-mode response increases toward larger $\lambda_{\rm GW}/L$, similar to the familiar $1/f$-type roll-off of a Michelson-like interferometric response in this regime. In contrast, the differential response remains suppressed due to the cancellation of the leading long-wavelength contribution in the folded-loop geometry.

The resonant peaks themselves follow an overall envelope that decreases as $1/f$ toward shorter GW wavelengths (higher resonance order). This behavior reflects the fact that, although coherent accumulation persists at the discrete resonance conditions, an increasing fraction of the GW-induced phase cancels along the optical path as the GW wavelength becomes shorter than the loop scale.

Notably, at the resonant frequencies, the GW response of the folded-loop interferometer is purely differential in character: the resonant enhancement of the differential (CW–CCW) response is accompanied by a common (CW+CCW) response that is strongly suppressed, in fact, vanishing in the limit $|\boldsymbol{\delta}|\to 0$. This behavior follows from the symmetry of the folded-loop geometry and the opposite orientation of the GW-induced phase accumulation for CW and CCW propagation. While differential readout is primarily employed to suppress common-mode noise, the absence of GW-induced resonances in the common channel at the differential-mode resonances provides a particularly clean spectral signature. In contrast to geometries where resonant features appear simultaneously in both readout channels, the folded-loop response localizes the resonant GW signal entirely in the differential channel.

For comparison, it is instructive to consider other interferometric geometries employing kinked optical paths that have been proposed in the literature. The L-shaped (Fox–Smith–type) resonant interferometer introduced in Ref.~\cite{Zhang:2022yab} achieves resonant enhancement by folding the optical path within a single kinked arm. In that configuration, however, the circulating optical fields propagate along the same physical path, and the readout does not involve two independently propagating counter-propagating beams. A closer point of comparison is provided by hybrid or double-L geometries, recently discussed in Ref.~\cite{Guo:2024cgz}, which combine folded optical paths with a Michelson-like differential readout. For the triangular ($60^\circ$) tunnel configuration compatible with the ET layout considered here, we find that the peak differential response at resonance is reduced by a factor of two compared to the folded-loop geometry, and that resonant enhancement occurs simultaneously in both the differential- and common-mode responses. A quantitative comparison of the corresponding response functions is presented in Appendix~\ref{app:comparison_geometries}.

\subsection{Projected sensitivity}

To translate the response functions discussed above into an estimate of the achievable strain sensitivity, one would in principle need to construct a complete noise budget for the folded-loop interferometer, including optical, mechanical, and quantum noise sources, as well as their coupling to the relevant readout modes. Developing such a full model for a novel detector geometry is a substantial undertaking and lies beyond the scope of the present work.

Instead, we adopt a calibrated estimate based on a comparison with the ET, whose projected strain sensitivity is well studied. The basic idea is to factorize the strain sensitivity schematically as
\begin{equation}
S_h^{1/2}(\omega_{\rm GW}) \sim 
\frac{S^{1/2}_{\mathrm{noise}}(\omega_{\rm GW})}{R_{\mathrm{rms}}(\omega_{\rm GW})}\,,
\end{equation}
where $R_\mathrm{rms}(\omega_\mathrm{GW})$ denotes the idealized detector response\footnote{The response functions computed in this work are most naturally interpreted as an idealized response: they capture the intrinsic, geometry-driven phase response of the interferometer, including retardation and resonant accumulation effects, but abstract from a detailed modeling of optical and technical noise sources. In this sense, they play a role analogous to, but not identical with, the purely geometrical antenna patterns often used in the analysis of conventional Fabry--Pérot Michelson interferometers.} discussed above and $S^{1/2}_{\mathrm{noise}}$ is the (effective) amplitude spectral density of readout phase noise in the relevant channel.

\begin{figure}[t]
\centering
\includegraphics[width=0.494\textwidth,trim=0.33cm 0.25cm 0.0cm 0.0cm, clip]{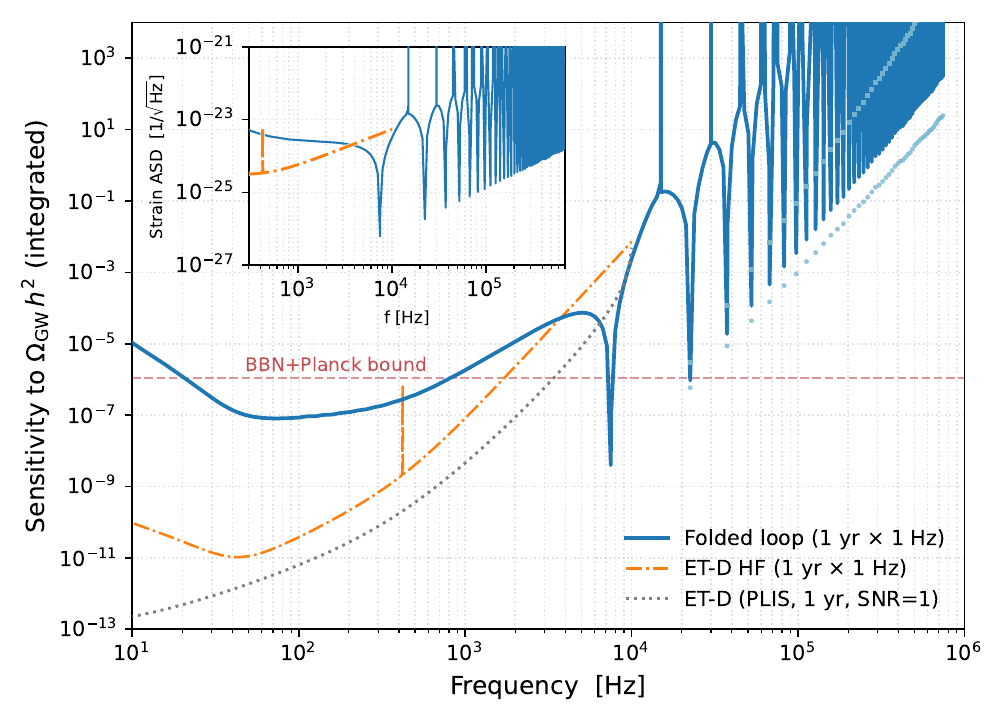}
\caption{Projected sensitivity (calibrated estimate) of a folded-loop interferometer with ET-like dimensions ($L=10,\mathrm{km}$, $60^\circ$ opening angle) and finesse $\mathcal{F}=500$ (blue solid curves). The main panel shows the one-year integrated sensitivity to a stochastic GW background, while the inset displays the corresponding instantaneous strain response $S_h^{1/2}(f)$.
The orange dot-dashed curve indicates the ET high-frequency (ET-HF) design sensitivity~\cite{Hild:2010id,Kaiser:2020tlg}, and the gray dotted curve shows the ET power-law–integrated sensitivity (PLIS)~\cite{Schmitz:2020syl} for reference. The red dashed line denotes the cosmological upper bound from BBN and Planck data~\cite{Caprini:2018mtu,Yeh:2022heq}.
Light-blue markers indicate the resonance minima obtained under alternative assumptions for the calibration beyond $10\,\mathrm{kHz}$ (constant and shot-noise–dominated extrapolations), as discussed in the text.}
\label{fig:sensitivity}
\end{figure}

Assuming that a folded-loop interferometer can be realized with optical performance comparable to that of ET, we estimate its strain sensitivity by calibrating against the ET response. Concretely, we compute the idealized response for both ET and the folded-loop geometry and rescale the known ET sensitivity accordingly. While this procedure is necessarily approximate, it provides a reasonable first estimate of the sensitivity reach implied by the response functions derived above.

Figure~\ref{fig:sensitivity} shows the resulting sensitivity estimate for a folded-loop interferometer with ET-like dimensions, taking $L=10\,\mathrm{km}$ and a $60^\circ$ opening angle compatible with the triangular ET tunnel layout. The calculation assumes a cavity finesse $\mathcal{F}=500$, corresponding to an effective number of coherent round trips $\langle n_{\rm rt}\rangle \simeq 160$. The main panel shows the one-year integrated sensitivity to an isotropic stochastic background, while the inset displays the corresponding instantaneous strain response $S_h^{1/2}(f)$.

The sensitivity curve exhibits pronounced narrowband features associated with the resonant enhancement of the differential response. In the folded-loop geometry, the fundamental resonance occurs when the GW wavelength is commensurate with the loop scale,
$\lambda_{\rm GW} \simeq 4L$, corresponding to $f \simeq c/(4L)\approx 7.5\,\mathrm{kHz}$ for $L=10\,\mathrm{km}$, followed by a sequence of higher resonances at $f \propto (2\ell+1)$.
At these resonant frequencies, the folded loop reaches its maximal response in the differential channel, and the resulting sensitivity dips form a characteristic comb-like spectral signature.

In terms of scaling, the integrated sensitivity to $\Omega_{\rm GW}$ improves as $T^{-1/2}$ with observation time, as expected for stationary noise, and the resonant depth scales approximately as $\langle n_{\rm rt}\rangle^{-2}$ because the phase response grows linearly with the effective number of coherent traversals. Increasing the finesse therefore enhances the narrowband reach and deepens the resonance minima, at the expense of bandwidth set by the cavity storage time. Varying the baseline length $L$ shifts the resonance comb in frequency, with larger $L$ moving all resonances to lower frequencies while also increasing the phase response per traversal. Notably, shorter baselines do not yield a parametric gain in the envelope of the resonant extrema at high frequencies. The reduced geometric phase accumulation of a shorter loop exactly compensates the upward shift of its resonance comb, so that the sensitivity reached at the lowest resonances of a short-baseline interferometer is parametrically comparable to that of higher-order resonances of a longer-baseline device.

A practical limitation of the ET-based calibration is that published ET strain sensitivities (and hence our calibration input) are available only up to a frequency of $10\,\mathrm{kHz}$. To extend the sensitivity estimate into the higher-frequency regime relevant for the subsequent folded-loop resonances, we extrapolate the calibration beyond $10\,\mathrm{kHz}$. Our baseline choice (blue curves in Fig.~\ref{fig:sensitivity}) continues the calibration with the effective log--log slope extracted from the highest-frequency bins of the ET calibration data, which lies between a constant continuation and the asymptotic shot-noise-limited behavior. Since shot noise ultimately dominates the effective noise budget at sufficiently high frequencies, one expects the noise term to scale as $S^{1/2}_{\mathrm{noise}}(f)\propto f$ in this regime, which in turn implies an asymptotic high-frequency envelope in $\Omega_{\rm GW}h^2$ must eventually approach $\propto f^{7}$, whereas a constant continuation of the calibration would give $\propto f^{5}$. To illustrate the impact of this uncertainty without introducing a detailed noise model for the folded-loop optics, we indicate in Fig.~\ref{fig:sensitivity} (light blue markers) the resonance minima obtained under both alternative continuations (constant and shot-noise–dominated), matched at $10\,\mathrm{kHz}$.\footnote{Unlike broadband detectors optimized for the $10$--$10^3\,\mathrm{Hz}$ band, a folded-loop interferometer targeting narrow high-frequency resonances could, in principle, rebalance quantum noise contributions to favor the resonant band. Assessing such detector-specific optimizations lies beyond the scope of the present work, but this consideration motivates treating the high-frequency extrapolation as indicative rather than definitive.
}

With these assumptions, the folded-loop geometry reaches below the cosmological upper bound inferred from BBN and Planck over the first resonances, corresponding to frequencies of order $10\,\mathrm{kHz}$ and above, and hence probes stochastic backgrounds in a regime that is largely inaccessible to conventional kilometer-scale Michelson interferometers. This reach is particularly relevant for early-Universe scenarios that generate high-frequency backgrounds associated with energy scales far above those probed by the LIGO/Virgo/KAGRA band.

Taken together, the folded-loop response exhibits a narrowband, comb-like resonance structure whose frequencies are fixed entirely by geometry and whose appearance is intrinsically differential. This behavior is a direct consequence of the transverse-traceless nature of GWs and the symmetry of the folded optical path, rather than of detector-specific noise properties.
As a result, the resonant signal is difficult to mimic by known instrumental or environmental noise sources, which typically lack both the required geometric frequency locking and the correlated cancellation between readout channels. In this sense, the folded-loop response provides a characteristic ``smoking-gun'' signature for high-frequency GWs and does not rely on cross-correlation with independent detectors to establish a GW origin.

\subsection{Feasibility and challenges}

In practice, the folded-loop interferometer would be operated as an optical resonator, with the total optical path length actively stabilized to an integer multiple of the laser wavelength. Throughout this work, we implicitly assume optical parameters comparable to those of the ET, for which a laser wavelength of order $10^3\,\mathrm{nm}$ is envisaged. Such stabilization schemes are standard in GW interferometry and rely on active control of mirror positions to maintain resonance.

Optical stability at non-normal incidence is likewise well established. High-finesse triangular ring cavities with incidence angles in the range $30^\circ$--$60^\circ$ are routinely operated as input mode cleaners in existing detectors, including GEO600 and Advanced LIGO\@. For example, the Advanced LIGO input mode cleaner has a round-trip length of $32.9\,\mathrm{m}$ and a finesse of order $500$, demonstrating that low-loss, high-finesse operation in folded geometries is feasible with current technology~\cite{LIGOScientific:2014pky}.

The folded-loop geometry considered here introduces additional challenges, including maintaining high finesse in the presence of non-normal reflections at multiple mirrors, controlling potential backscattering effects arising from the proximity of mirrors TM1 and TM3,\footnote{The most direct handle on potential backscattering is the separation between TM1 and TM3, which can be adjusted to reduce overlap between counter-propagating beams while maintaining a small enclosed area to control rotational non-reciprocity. If these requirements were to compete in practice, potential backscattering effects associated with the near-normal reflections at TM2 and TM4 could also be mitigated by implementing each of these direction changes via pairs of steeper reflections rather than single shallow-angle mirrors, thereby introducing a small spatial offset between incoming and outgoing beams and suppressing coherent back-reflection without significantly enlarging the enclosed area.} and achieving the geometric alignment required to suppress residual rotational non-reciprocity. While these requirements differ in detail from those of conventional Michelson interferometers, they act predominantly on local optical elements rather than on kilometer-scale baselines.

At the same time, the high degree of path commonality between the CW and CCW beams implies that many technical disturbances are naturally imprinted as common-mode fluctuations and are therefore suppressed by differential readout. While a quantitative assessment of such effects requires a dedicated noise model, this structural feature may provide additional robustness against certain technical noise sources compared to interferometers with spatially separated arms.

Overall, the experimental requirements of the folded-loop concept appear compatible with the technological scope of next-generation interferometers such as ET\@. A detailed optical and noise model tailored to this geometry would be a natural next step, but lies beyond the scope of the present, concept-driven study.

\section{Conclusions}
\label{sec:conlcusion}

We have presented a new resonant interferometric concept for probing high-frequency GWs based on coherent phase accumulation in closed optical loops. The key idea is to exploit geometric resonances in which the GW-induced phase shift adds coherently over many loop traversals, producing sharp, narrowband sensitivity enhancements. The resulting resonance comb constitutes a distinctive and predictable signature that directly reflects the transverse--traceless nature of GWs and is difficult to mimic by instrumental or environmental noise.

Focusing on a folded-loop geometry, we have shown how this phase-accumulation mechanism can be realized in a form compatible with large-scale terrestrial detectors. By folding the loop to suppress rotational non-reciprocity, the design avoids the Sagnac limitation that obstructs simple ring resonators on Earth, while retaining the resonant response central to the concept. A salient feature of the folded loop is that the resonant GW signal appears exclusively in the differential (CW--CCW) readout channel, with the common-mode response vanishing at the same frequencies. This purely differential resonance structure provides a particularly clean spectral signature and, in principle, allows a stochastic background to be identified without relying on cross-correlation between multiple detectors.

Using an ET-scale implementation as a benchmark, with loop length $L=10\,\mathrm{km}$ and a realistic finesse $\mathcal{F}=500$ (corresponding to $\langle n_{\rm rt}\rangle\simeq160$), we have presented sensitivity projections based on a calibrated comparison with the ET\@. For one year of observation, the folded-loop interferometer reaches below the cosmological bounds from BBN and CMB data at its first resonances, probing stochastic backgrounds up to frequencies of order $20$--$30\,\mathrm{kHz}$. In cosmological terms, this corresponds to processes in the early Universe at temperatures $\gtrsim10^{9}\,\mathrm{GeV}$, far beyond the reach of CMB or conventional ground-based interferometers.

The sensitivity exhibits a characteristic resonance comb whose positions are fixed by geometry, while the envelope reflects the interplay between coherent phase accumulation and retardation along the loop. Increasing the effective number of coherent round trips, $\langle n_{\rm rt}\rangle$, enhances the resonant response linearly such that the sensitivity to $\Omega_\mathrm{GW} h^2$ scales as $\langle n_{\rm rt}\rangle^{-2}$, while changing the loop scale shifts the resonance pattern in frequency. These scaling properties make ET-scale infrastructures a natural benchmark for high-frequency searches, while leaving open the possibility of optimization toward specific frequency bands.

The principal challenges of the folded-loop concept are technical rather than conceptual, including high-finesse operation at non-normal incidence, control of backscattering in folded sections, and precise geometric alignment to suppress residual non-reciprocity. All of these challenges act on local optical elements and build directly on techniques already demonstrated in existing GW detectors. A full optical and noise model tailored to this geometry is a natural next step, but lies beyond the scope of the present, concept-driven study.

Overall, resonant folded-loop interferometers offer a practical and distinctive pathway toward exploring an uncharted region of the GW spectrum. Their narrowband, geometry-driven response not only enhances sensitivity at targeted frequencies, but also provides an unmistakable experimental hallmark of GWs at high frequencies. We hope that this work will stimulate further investigation of resonant loop geometries within the GW detector community.

\section*{Acknowledgments}
I thank Valerie Domcke, Mathias Garny, and Jörn Kersten for valuable comments on the manuscript, and Julien Lesgourgues for stimulating discussions. 
I am particularly grateful to Harald L\"uck for pointing out the relevance of rotational non-reciprocity, which proved crucial for the development of this work.

\appendix

\section{Closed-form expressions for phase accumulation}
\label{app:closed_form_sums}

In this appendix we collect the explicit definitions and closed-form expressions
for the sums introduced in Sec.~\ref{sec:general_formalism}.
In particular, we provide the definitions of $\beta_s^{(m)}$ and show how
Eq.~\eqref{eq:DeltaPhi_n} and Eq.~\eqref{eq:linear_superposition}
can be evaluated analytically.

\subsection{Definition of $\alpha_s$ and $\beta_s^{(m)}$}

For a plane GW with angular frequency $\omega_{\rm GW}$
propagating in direction $\hat k$, we define
\begin{equation}
\alpha_s
\equiv
\omega_{\rm GW}\bigl(1-\hat k\!\cdot\!\hat e_s\bigr),
\label{eq:alpha_def_app}
\end{equation}
as introduced in Sec.~\ref{sec:general_formalism}.

The phase offset $\beta_s^{(m)}$ appearing in
Eq.~\eqref{eq:segment_phase} encodes both the segment position
within the loop and the traversal index $m$.
Using the parametrization of the photon trajectory
\begin{equation}
t_s^{(m)}(x)
=
m L_{\rm rt} + \Sigma_s + x,
\end{equation}
with
\begin{equation}
\Sigma_s
\equiv
\sum_{r=1}^{s-1} L_r,
\qquad
L_{\rm rt}
=
\sum_{r=1}^{N_{\rm seg}} L_r,
\end{equation}
one finds
\begin{equation}
\beta_s^{(m)}
=
\omega_{\rm GW}
\Bigl(
m L_{\rm rt}
+
\Sigma_s
-
\hat k\!\cdot\!\mathbf{x}_{s,0}
\Bigr)
+
\phi_0 .
\label{eq:beta_def_app}
\end{equation}

The quantities $\alpha_s$ and $\beta_s^{(m)}$
completely determine the GW phase along segment $s$
during traversal $m$.

\subsection{Closed-form evaluation of $\Delta\Phi(n)$}

The total phase accumulated after $n$ completed traversals is
\begin{equation}
\Delta\Phi(n)
=
\sum_{m=0}^{n-1}
\sum_{s\in\mathcal{P}}
\delta\Phi_s^{(m)},
\label{eq:DeltaPhi_n_app}
\end{equation}
with $\delta\Phi_s^{(m)}$ given in Eq.~\eqref{eq:segment_phase}.

From Eq.~\eqref{eq:beta_def_app}, the traversal dependence of each segment contribution enters exclusively through the term
\begin{equation}
\beta_s^{(m)}
=
\beta_s^{(0)}
+
m\,\omega_{\rm GW} L_{\rm rt},
\end{equation}
so that successive round trips differ by a fixed phase advance
\begin{equation}
\delta\phi
\equiv
\omega_{\rm GW} L_{\rm rt}.
\label{eq:delta_phi_def}
\end{equation}

It is therefore convenient to introduce a complex representation of the
segment contribution,
\begin{equation}
\delta\Phi_s^{(m)}
=
\Re\!\left[
\widetilde{\Phi}_s
\,\E^{\I m\delta\phi}
\right],
\end{equation}
where the complex segment amplitude is
\begin{equation}
\widetilde{\Phi}_s
=
\frac{\pi L_s}{\lambda_{\rm laser}}
\,
\mathrm{sinc}\!\left(\frac{\alpha_s L_s}{2}\right)
\Bigl(
h_+ P^+_s
-
\I h_\times P^\times_s
\Bigr)
\E^{\I\left(\beta_s^{(0)}+\frac{\alpha_s L_s}{2}\right)} .
\label{eq:Phi_s_tilde}
\end{equation}
The polarization projection factors are defined as
\begin{equation}
P_s^{+}
\equiv
e^{+}_{ij}(\hat k,\psi)\,
\hat e_s^i \hat e_s^j,
\qquad
P_s^{\times}
\equiv
e^{\times}_{ij}(\hat k,\psi)\,
\hat e_s^i \hat e_s^j,
\label{eq:Ps_def_app}
\end{equation}
where $e^{+,\times}_{ij}(\hat k,\psi)$ denote the standard
transverse-traceless polarization tensors for a plane GW
propagating in direction $\hat k$ with polarization angle $\psi$.
Summing over all segments yields the single-traversal complex loop amplitude
\begin{equation}
\widetilde{\Phi}_{\rm loop}
=
\sum_{s\in\mathcal{P}}
\widetilde{\Phi}_s .
\label{eq:Phi_loop_tilde}
\end{equation}

The finite traversal sum in Eq.~\eqref{eq:DeltaPhi_n_app}
can now be evaluated as a geometric series,
\begin{equation}
\Delta\Phi(n)
=
\Re\!\left[
\widetilde{\Phi}_{\rm loop}
\,
\frac{1-\E^{\I n\delta\phi}}
{1-\E^{\I\delta\phi}}
\right].
\label{eq:DeltaPhi_n_closed}
\end{equation}

Equation~\eqref{eq:DeltaPhi_n_closed} makes explicit the resonant enhancement
when $\delta\phi \simeq (2\ell+1)\pi$.

\subsection{Closed-form cavity response}

In a resonant loop cavity, the circulating phase is given by
\begin{equation}
\Delta\Phi_{\rm cav}
=
\sum_{n=0}^{\infty}
w_n\,\Delta\Phi(n),
\label{eq:cavity_sum_app}
\end{equation}
with $w_n \propto r^n$ for an effective round-trip amplitude reflectivity $r$.

Using Eq.~\eqref{eq:DeltaPhi_n_closed} and performing the infinite sum over $n$
yields
\begin{equation}
\Delta\Phi_{\rm cav}
=
\Re\!\left[
\widetilde{\Phi}_{\rm loop}
\,
\frac{1}{1-r \E^{\I\delta\phi}}
\right],
\label{eq:cavity_closed_app}
\end{equation}
valid for $|r|<1$.

This expression underlies the numerical response functions shown in
Sec.~\ref{sec:response}.  The effective number of coherent round trips
is
\begin{equation}
\langle n_{\rm rt}\rangle
=
\sum_{n=0}^{\infty} r^n
=
\frac{1}{1-r}
\simeq
\frac{\mathcal{F}}{\pi},
\end{equation}
for a high-finesse cavity.

\section{Folded-loop geometry and full response}
\label{app:folded_geometry}

This appendix collects additional details on the folded-loop interferometer
geometry and shows response functions over an extended frequency range beyond
the region emphasized in the main text.

\subsection{Geometry specification}
\label{app:folded_geometry_spec}

We model each optical path as a sequence of straight segments connecting idealized
test-mass positions (``vertices''). For any directed segment from vertex
$i$ to vertex $j$ the unit tangent vector is
\begin{equation}
\hat e_{i\to j}
\;=\;
\frac{\mathbf{x}_j-\mathbf{x}_i}{|\mathbf{x}_j-\mathbf{x}_i|},
\qquad
L_{i\to j}=|\mathbf{x}_j-\mathbf{x}_i|.
\label{eq:evec_from_coords}
\end{equation}
The segment starting point is $\mathbf{x}_{s,0}=\mathbf{x}_i$ and the intra-loop
offset $\Sigma_s$ follows from the ordered sum of previous segment lengths,
cf.\ Eq.~\eqref{eq:beta_def_app}.

For the ET-compatible ($60^\circ$) folded-loop realization used throughout this
work we choose coordinates (in the plane $z=0$)
\begin{align}
\mathbf{x}_{\rm TM1} &= (0,0,0), \nonumber\\
\mathbf{x}_{\rm TM2} &= (L,0,0), \nonumber\\
\mathbf{x}_{\rm TM4} &= \Bigl(\tfrac{L}{2}, \tfrac{\sqrt{3}}{2}L,0\Bigr), \nonumber\\
\mathbf{x}_{\rm TM3} &= (0,0,0) + \boldsymbol{\delta},
\qquad |\boldsymbol{\delta}|\ll L.
\label{eq:folded_coords_app}
\end{align}
In the response computation we neglect the small separation between TM1 and TM3
and set $\mathbf{x}_{\rm TM3}\to \mathbf{x}_{\rm TM1}$ in the retarded GW phase
offsets (i.e.\ in $\beta_s^{(m)}$), while keeping the folded topology of the optical
path. (In an actual realization, the displacement vector $\boldsymbol{\delta}$ would be adjusted parallel to the Earth's rotation axis, $\boldsymbol{\delta}\parallel \boldsymbol{\Omega}_\oplus$.)

With these vertices, the two directed paths are:
\begin{align}
\mathcal{P}_{\rm CW}:&\quad
{\rm TM1}\to{\rm TM4}\to{\rm TM3}\to{\rm TM2}\to{\rm TM1},
\label{eq:path_CW_folded_app}\\
\mathcal{P}_{\rm CCW}:&\quad
{\rm TM1}\to{\rm TM2}\to{\rm TM3}\to{\rm TM4}\to{\rm TM1}.
\label{eq:path_CCW_folded_app}
\end{align}
All segment unit tangents $\hat e_s$ and lengths $L_s$ follow directly from
Eq.~\eqref{eq:evec_from_coords}.

\subsection{Full-frequency response}
\label{app:folded_full_response}

\begin{figure*}[t]
\centering
\includegraphics[width=0.48\textwidth]{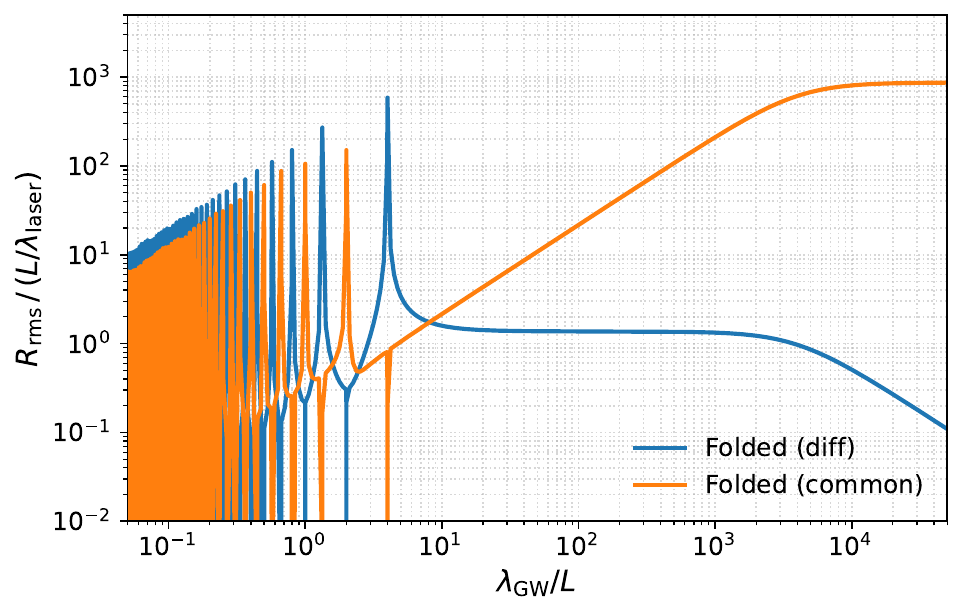}
\hfill
\includegraphics[width=0.48\textwidth]{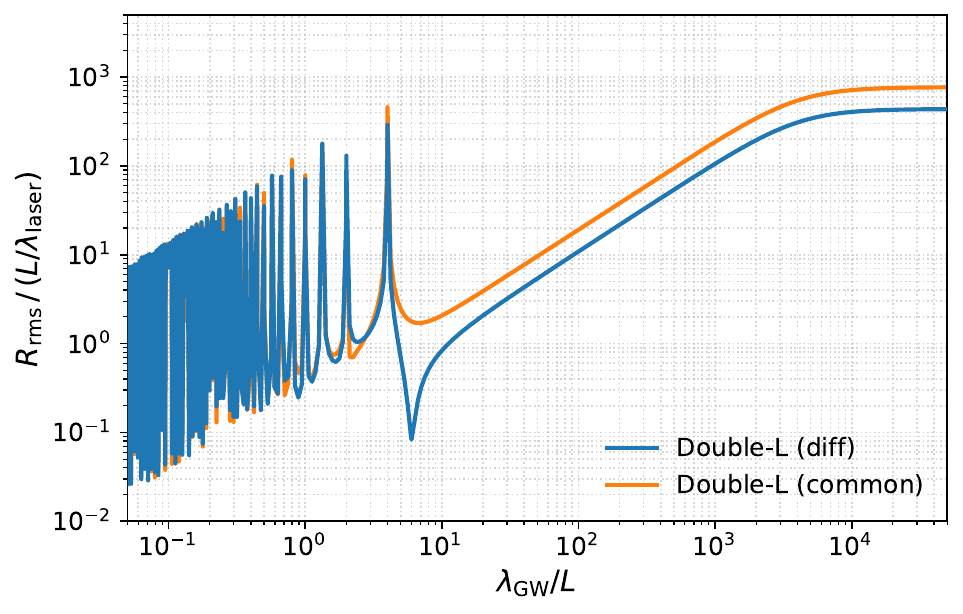}
\caption{Full-frequency RMS response functions. \textbf{Left:} folded-loop geometry
(App.~\ref{app:folded_geometry}). \textbf{Right:} double-L (hybrid) geometry
(App.~\ref{app:comparison_geometries}) adapted to an ET-compatible ($60^\circ$)
layout. Both panels show differential and common readout modes. Note that for readability, in the right panel the differential curve is drawn on top of the common-mode curve.}
\label{fig:response_full_compare}
\end{figure*}

Figure~\ref{fig:response_full_compare} (left panel) shows the RMS response of the
folded-loop interferometer over an extended frequency range, including the
long-wavelength regime beyond the $1/f$ roll-off discussed in the main text.
The plot illustrates (i) the suppression of the differential response in the
deep long-wavelength limit, (ii) the emergence of discrete geometric resonances,
and (iii) the decreasing envelope of resonance peaks toward higher resonance order.

\section{Comparison with the double-L geometry}
\label{app:comparison_geometries}

For comparison, we consider the hybrid (``double-L'') interferometer proposed in
Ref.~\cite{Guo:2024cgz}, adapted to a triangular ($60^\circ$) facility layout compatible
with ET\@. The purpose of this appendix is not to provide a full optical design, but to
specify the idealized piecewise-straight paths used in our response calculation and to
contrast their response features with those of the folded loop.

\subsection{Geometry and readout}

\begin{figure}[t]
\centering
\includegraphics[width=0.3\textwidth]{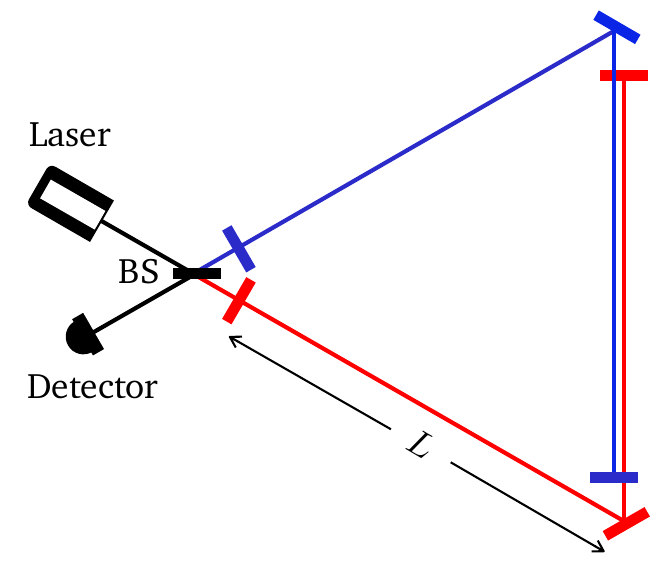}
\caption{Schematic of the double-L (hybrid) interferometer. A beamsplitter (BS) injects light into two
kinked arms; each arm contains a change of direction within the resonator. The readout is Michelson-like,
formed by recombining the arm fields at the BS.}
\label{fig:doubleL_geometry}
\end{figure}

A schematic layout is shown in Fig.~\ref{fig:doubleL_geometry}.
Light from the laser is split at a beamsplitter (BS) into two arms, and recombined at the
BS before readout at the photodetector. Each arm is implemented as a folded (kinked)
Fabry-P\'erot-like resonator, i.e.\ within each arm the circulating field undergoes one
change of direction between the near and far optics.

In a realistic implementation, the first steering optics of each arm (the ``near'' test
masses) are placed close to the beamsplitter compared to the long scale $L$ of the folded
segments. In the response calculation we idealize this by neglecting the short BS-to-near-mirror
offsets, i.e.\ we treat the long folded segments as starting effectively at a common vertex.
This approximation does not affect the geometric GW response in the frequency band of interest,
which is controlled by retardation along the long segments of length $\sim L$.

For the ET-like ($60^\circ$) configuration considered here, the two arms start in directions
separated by $60^\circ$, and the kink in each arm is chosen such that the two long segments in
that arm have (approximately) equal length $L$ and fit within the triangular tunnel geometry.
The differential readout is the Michelson-like combination between the two arms, while the
corresponding common channel is obtained by the sum of the two arm signals.

\bigskip

\subsection{Response comparison}

Figure~\ref{fig:response_full_compare} compares the RMS response functions of the folded-loop and double-L
geometries. For the triangular ($60^\circ$) configuration considered here, the peak differential response
of the double-L geometry is reduced by a factor of two relative to the folded loop, and resonant enhancement
occurs simultaneously in both differential and common channels.

This contrasts with the folded-loop geometry, where the resonant enhancement is confined to the differential
(CW--CCW) response while the common-mode GW response vanishes at the same resonance frequencies.

\bibliographystyle{bjstyle}
\bibliography{bib}

\end{document}